\documentclass[12pt]{iopart}
\usepackage{color}
\usepackage{units}
\usepackage{amssymb}
\usepackage{graphicx}
\usepackage[unicode=true,
 bookmarks=true,bookmarksnumbered=false,bookmarksopen=false,
 breaklinks=false,pdfborder={0 0 1},backref=false,colorlinks=true]
 {hyperref}

\begin{document}

\title{Sustained propagation and control of topological excitations in polariton superfluid}

\author{Simon Pigeon$^{1,2}$, Alberto Bramati$^1$}
\address{$^{1}$Laboratoire Kastler Brossel, UPMC-Sorbonne Universit\'es, CNRS, ENS-PSL
Research University, Coll\`ege de France, 4 place Jussieu Case 74, F-75005
Paris, France}
\address{$^{2}$Centre for Theoretical Atomic, Molecular and Optical Physics, School
of Mathematics and Physics, Queen's University Belfast,
Belfast BT7 1NN, United Kingdom}
\ead{simon.pigeon@lkb.upmc.fr}

\begin{abstract}
We present a simple method to compensate for losses in a polariton superfluid. Based on a weak support field, it allows for an extended propagation of a resonantly driven polariton superfluid at a minimal energetic cost. Moreover, this setup based on optical bistability, leads to a significant release of the phase constraint imposed by the resonant driving. This release, together with the macroscopic polariton propagation,  offers a unique opportunity to study the hydrodynamics of topological excitations of polariton superfluids such as quantized vortices and dark solitons. We numerically study how the coherent field supporting the superfluid flow interacts with the vortices and how it can be used to control them. Interestingly, we show that standard hydrodynamics does not apply for this driven-dissipative fluid and new behaviours are identified. 
\end{abstract}

\section{Introduction}

The control and manipulation of quantum states is a challenging
goal in modern physics. Among the recent developments, the
discovery of quantum states, such as Bose-Einstein condensates
and superfluids, in solid-state quasi-particles like exciton polaritons
opens new perspectives \cite{Kasprzak2006,Amo2009} for the study of the fundamental properties of quantum fluids on a semiconductor chip. In a semiconductor
microcavity with embedded quantum wells with an
excitonic transition resonant with the cavity mode, the strong coupling regime can be easily reached \cite{Weisbuch1992}, giving rise to system eigenmodes which
are an indistinguishable mix of excitons and cavity photons, called microcavity polaritons.
The unique balance between the dissipative nature of these quasi-particles
and their non-linear dynamics leads to the generation of a superfluid
of polaritons \cite{Carusotto2013}. 

A simple way to observe exciton polaritons is to quasi-resonantly drive
the system with a laser field. For homogenous excitation or a large
enough excitation spot, slightly blue-detuned with respect to the polariton energy, an optical bistable behaviour is observed \cite{Baas2004}. The associated
hysteresis cycle (c.f. Fig. \ref{fig:hyst}) delimits two different regimes: (i) a linear regime for a weak
driving field ($I<I_-$) and (ii) a highly non linear regime for a strong driving field ($I>I_+$), in which the system exhibits a superfluid behaviour.  In this regime, several specific
quantum fluids properties were predicted and observed recently in polaritonic
systems, e.g., frictionless propagation \cite{Amo2009}, vortex pair
and soliton generation \cite{Sanvitto2011,Amo2011,Pigeon2010,Nardin2011},
heralding a new way to study quantum hydrodynamics of superfluids. 

\begin{figure}
\centering
\includegraphics[width=0.65\textwidth]{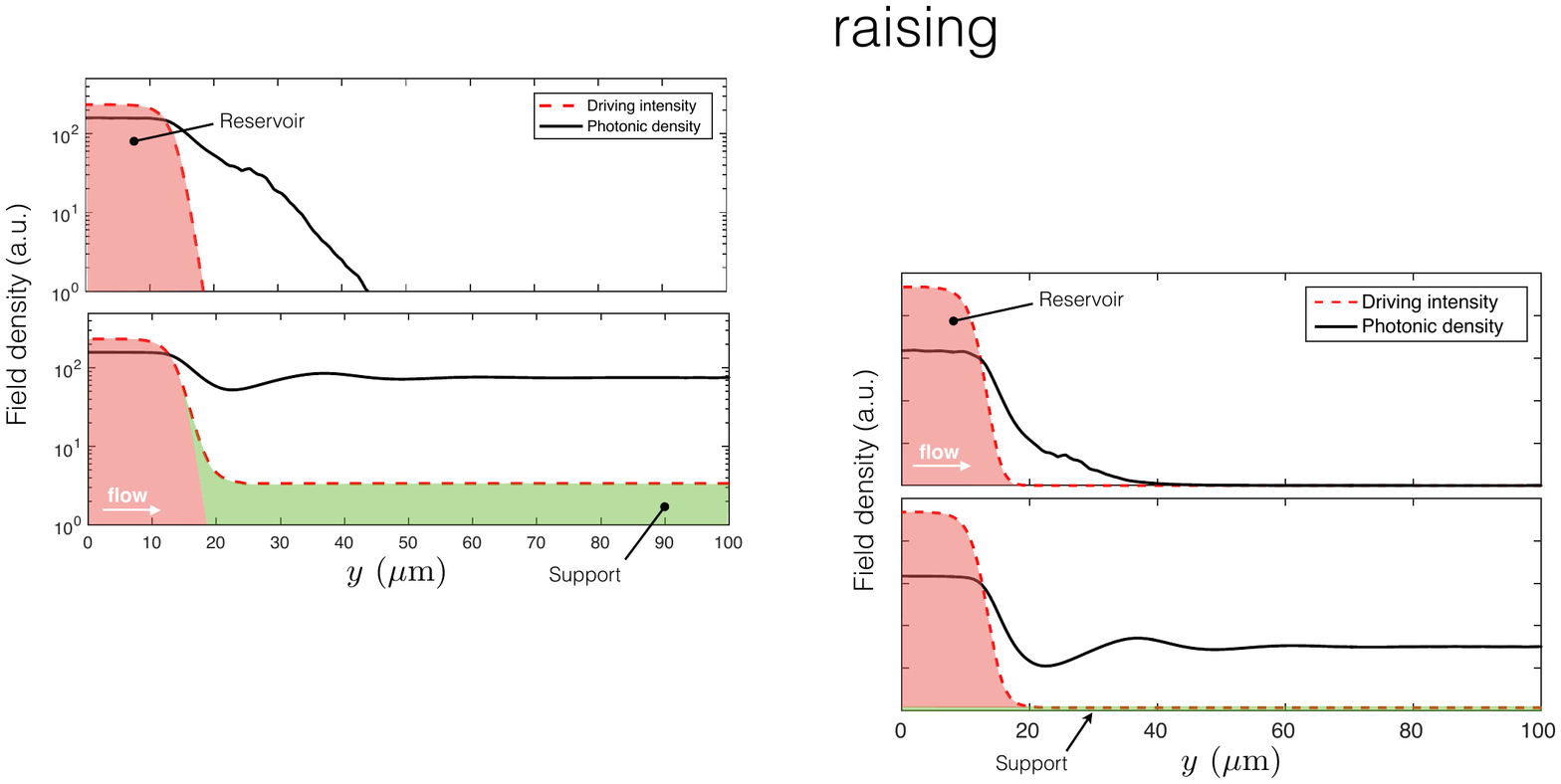}
\caption{Steady state photonic density (solid black curves) of a polariton superfluid
and corresponding driving profile (red dashed curves). The upper panel
represents the unsupported flow ($I/I_{+}=0$) and the lower panel
the supported flow ($I/I_{+}=0.08$, point A in Fig. \ref{fig:hyst}). Density in logarithmic scale.
The colored region delimits the pump (in red) and the support (in green)
driving. The parameters used are: $\hbar\omega_{C}(\mathbf{k}=\mathbf{0})=1602\unit{meV}$,
$\hbar\omega_{X}(0)=1600\unit{meV}$, $\hbar\gamma_X=\hbar\gamma_C=0.05\unit{meV}$,
$\hbar\Omega_R=2.5\unit{meV}$ and $\hbar g=0.01\unit{meV\mu m^2}$. The
driving field is such that $\Delta=\hbar\omega_{p}-\hbar\omega_{LP}(\mathbf{k}=\mathbf{0})=1\unit{meV}$
and $\vert\mathbf{k}_{p}\vert=(0,1)^{T}\unit{\mu m^{-1}}$.\label{fig:with_without}}
\end{figure}

In this article we study the
intermediate case in-between the linear and non-linear regime ($I_-<I<I_+$), the bistable regime. As we will show, the bistable
regime exhibits rich dynamical features, allowing for compensation
of the dissipation inherent to these fluids, and thus for extended propagation of the superfluid flow at low energetic
cost. Moreover, as shown in Ref.~\cite{Pigeon2010}, the resonant
driving field tends to lock the phase of the generated superfluid, inhibiting
the formation of topological excitations such as vortices or solitons.
Therefore, an engineered driving profile is required to observe
such fundamental excitations \cite{Sanvitto2011,Amo2011,Nardin2011}. Because of the inhomogeneous density profile of the fluid resulting from the engineered driving profile, this approach makes thorough hydrodynamic studies difficult. In this work
we show how with two driving fields, one to create the superfluid and one to support it, we can avoid the locking effect of the phase, allowing for the generation and propagation of vortices in a homogeneous superfluid. The combination of a macroscopic enhancement
of the propagation together with the absence of the phase locking effect
allows for detailed hydrodynamic analysis of vortex-pair dynamics
in superfluids. 
Interestingly, we show that a fine tuning of the properties of the topological excitations can be achieved by controlling the intensity of the support field. This flexibility allows to emphasize a specific hydrodynamic behaviour of these driven dissipative systems, which is at odd with the standard equilibrium hydrodynamics.

\section{Macroscopic enhancement of the superfluid propagation}

To explore the bistable regime we consider two driving fields
with same frequency $\omega_{p}$, in-plane
wave-vector $\mathbf{k}_{p}$ and circular polarisation. One driving field of high intensity ($I_r>I_+$) is localized and is used to create a polaritons superfluid. The second driving field, homogenous in time and space (infinitely-extended constant field), acts as a support field and is of weak intensity ($I_-<I_s<I_+$). 
The time-evolution in these conditions is described by a generalized Gross-Pitaevskii
equation \cite{Carusotto2013,Pigeon2010}: 
\begin{eqnarray}
 i\partial_{t}\left(\begin{array}{c}
\Psi_C(\mathbf{x},t)\\
\Psi_X(\mathbf{x},t)
\end{array}\right)&=&\hbar\left(\begin{array}{c}
F_s+F_r(\mathbf{x})\\
0
\end{array}\right)e^{-i(\mathbf{k}_{p}\mathbf{.x}-\omega_{p}t)}+\nonumber\\
&&\hbar\left(\begin{array}{cc}
\omega_{C}(\mathbf{k})+V(\mathbf{x})-\imath\frac{\gamma_C}{2} & \Omega_R\\
\Omega_R & \omega_{X}^{0}+g\vert\Psi_X(\mathbf{x},t)\vert^{2}-\imath\frac{\gamma_X}{2}
\end{array}\right)  \nonumber \label{ggp}\\
&&\quad\times\left(\begin{array}{c}
\Psi_C(\mathbf{x},t)\\
\Psi_X(\mathbf{x},t)
\end{array}\right),
\end{eqnarray}
where $\Psi_{C(X)}(\mathbf{x},t)$ represents the cavity (exciton)
field, $\hbar\omega_{C}(\mathbf{k})$ is the energy dispersion of the
cavity modes, and $\hbar\omega_{X}^{0}$ is the exciton
energy assuming no dispersion (infinite mass). $\gamma_{C(X)}$ is the decay rate of
the cavity (exciton) modes, and $g$ is the exciton--exciton interaction.
$V(\mathbf{x})$ is the photonic potential. The first term of Eq.(\ref{ggp})
refers to the driving field. $F_r(\mathbf{x})$ represents the driving amplitude profile
of the field necessary to create a polariton superfluid and $F_s$, the profile of the support driving ($I_s=\vert F_s \vert^2$). 

We illustrate the proposed method in Fig.~\ref{fig:with_without},
presenting  in logarithmic scale the steady-state photonic density of a polariton
superfluid, flowing from left to right (black curves) in a defect-free planar microcavity
($V=0$) for a total driving profile $F_r(\mathbf{x})+F_s$ represented by red dashed curves.
In the upper panel, only the driving $F_r(\mathbf{x})$, localised on the left,
acts on the system (i.e., $F_s=0$); it creates a  polariton superfluid
whose density falls exponentially outside the driven region due to the finite polariton lifetime. In this case, taking into account the current state of the art in microcavity fabrication which allows for polariton lifetime around 50 ps (in literature only one article \cite{Nelsen2013} reported for polariton longer lifetime, up to 100ps, for a very specific sample), the achievable propagation distances are limited to about $30\,\unit{\mu m}$. In strong contrast, the lower panel of Fig. \ref{fig:with_without} highlights, if the superfluid is
supported along its propagation, the polariton flow is maintained with a macroscopic density, without any decay. 
This suppression of the superfluid decay occurs even for weak support flow ($\vert F_r(0)\vert^2=I_r(0)\gg I_s\neq0$, shown in green). In Fig.~\ref{fig:with_without}, there are nearly two orders of magnitude between the intensity required to create the superfluid $I_r(0)$ and the intensity necessary to support its propagation $I_s$.
The modulations observed nearby the region where the superfluid is injected is due to the sharp profile of $F_r(\mathbf{x})$. 
The oscillations observed are dispersive shock waves \cite{Hoefer2006,Wan2007}. Without support, due to the strong decay experienced by the fluid, these oscillations are barely visible whereas with support, the shock waves are clearly visible, even thought rapidly attenuated.

In realistic experiments, the propagation distances will be mainly limited by the available power which fixes the maximum extension of the support. We anticipate that propagation distances of $100\,\unit{\mu m}$ will be easily obtained with standard setups; longer distances, up to $200\,\unit{\mu m}$ or more, will require specific arrangements (powerful lasers, mode shaping) but are still within reach. More importantly this strong enhancement of the propagation distances is obtained regardless of both the microcavity characteristics and the intrinsic polariton lifetime, relaxing the stringent fabrication requirements which are still very hard to fulfil, despite the recent technological improvements. 

\begin{figure}
\centering
\includegraphics[width=0.65\textwidth]{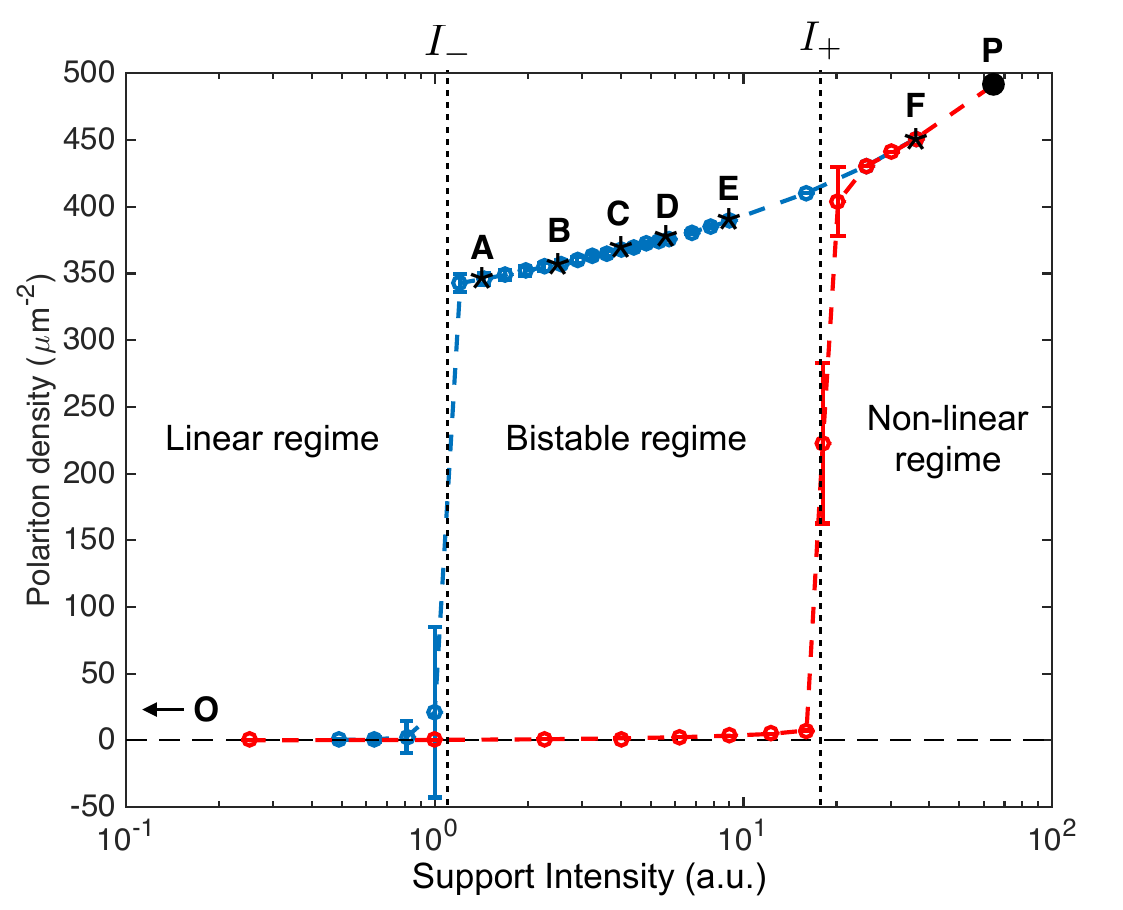}
\caption{Polariton density as a function of pump intensity. The red dashed
line with dots correspond to steady states obtained only in the presence
of the support driving ($F_r=0$), as the blue one represent the
case with jointly the pump and support driving on ($F_r\protect\neq0$).
The labeled black stars and dot refer to the panels of Fig. \ref{fig:panels}.
Other parameters are identical as in Fig. \ref{fig:with_without}.\label{fig:hyst}}
\end{figure}

This macroscopic enhancement of the superfluid propagation is a direct
consequence of the optical bistability. Bistability is typical for quasi-resonantly
driven Kerr non-linear media \cite{gibbs2012optical}. 
Many rich phenomena are related to this effect such as a large variety of solitonic solutions \cite{Egorov2009}. It is characterised by two distinct critical driving intensities, $I_{+}$ and $I_{-}$, ($I_{-}<I_{+}$) in-between which the system is bistable and can be either in linear regime or in a non-linear/superfluid regime. 
This behavior is clearly visible in Fig. \ref{fig:hyst} where
we plot the polariton density as a function of the support intensity. The density was obtained performing simulations similar to the one reported in Fig. 3 but omitting the potential barrier ($V=0$). The density was obtained through spatial averaging far enough from the region where the polariton superfluid is injected. Here the red curve corresponds to the support only whereas the blue curve to the support and pump.
If we trigger the superfluid regime with a local pump field,
the superfluid is maintained even with weak support intensity. The combination of pump and support fields has been previously explored in non-linear optical systems in a context of bright soliton written-erasure protocol \cite{Jenkins2009}, considering "light bullet" propagation \cite{Adrados2011} or optical circuit \cite{Liew2008}, also mixed with spin degree of freedom \cite{Amo2010} (not considered here).
In contrast, here we will use this scheme to explore the hydrodynamic
properties of the upper branch of the optical bistability. 

\begin{figure}[t!]
\centering
\includegraphics[width=1\textwidth]{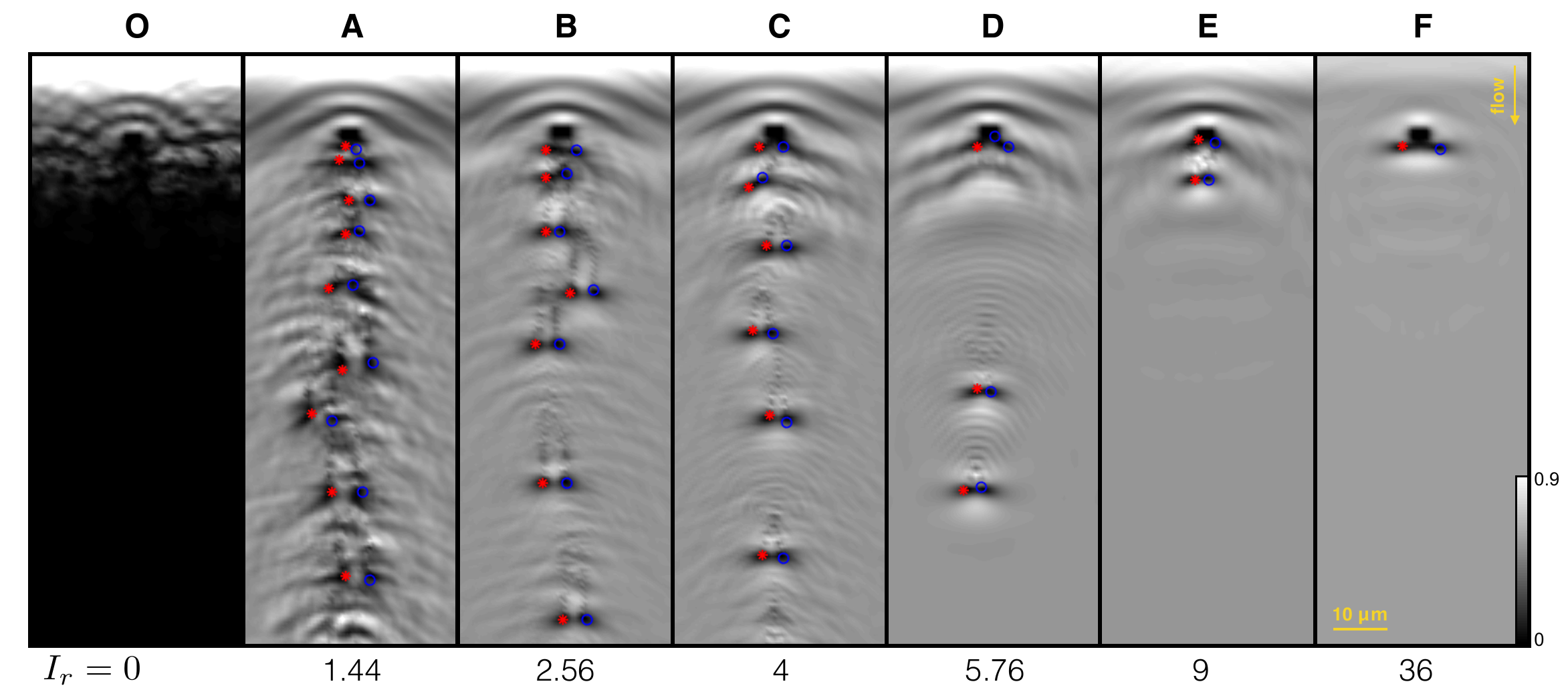}
\caption{Normalised density profile (snapshot) of polariton superfluid propagating through
a photonic defect for different support intensities (increasing from
left to right). The pump intensity localized above the photonic defect is fixed and corresponds to the black dot P in Fig. \ref{fig:hyst}. The label of each panel refers to a point on Fig. \ref{fig:hyst},
and the intensity of the support driving is indicated at
the bottom of each panel in arbitrary unit. The red dots and blue circles indicate,
respectively, vortices and antivortices. Other parameters are identical
to in Fig. \ref{fig:with_without}.\label{fig:panels}}
\end{figure}

\section{Sustained superfluid propagation and hydrodynamics}

\begin{figure}[t!]
\centering
\includegraphics[width=1\textwidth]{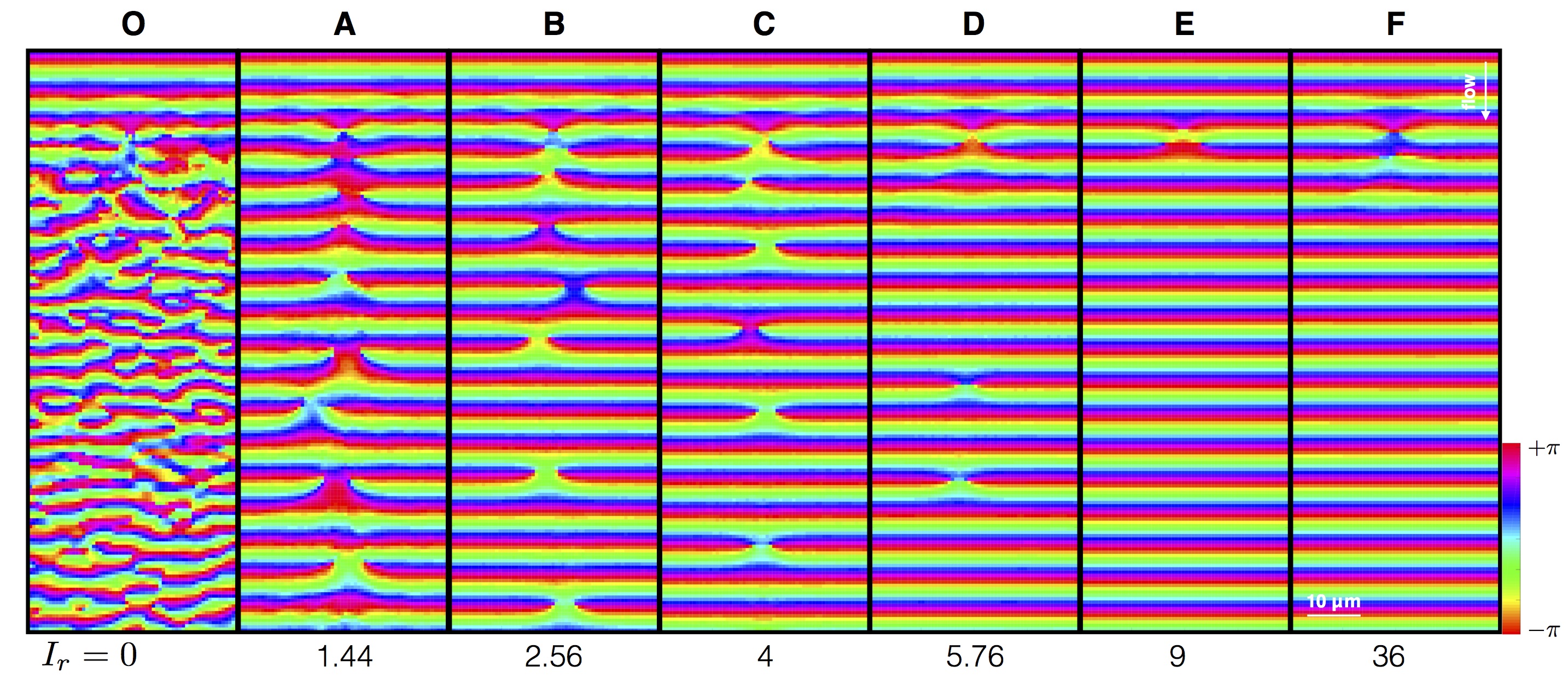}
\caption{Phase profile (snapshot) of polariton superfluid corresponding to the different panels of Fig. \ref{fig:panels} for increasing support intensity from
left to right. \label{fig:panels2}}
\end{figure}

In order to probe the superfluidity and to reveal its characteristics,
we place a large localized photonic potential barrier in the stream of the
fluid just below the pump region ($V\neq0$). A similar scheme without support
has been used to identify the different hydrodynamical regimes of a polariton superfluid~\cite{Pigeon2010}. Having just the pump above the potential barrier allows for a total release
of the locking of the phase due to resonant driving. Depending
on the flow speed and speed of sound ($c_s=\sqrt{mgn/\hbar}$ with
$m$ the polariton effective mass and $n$ the polariton density), different
hydrodynamical regimes can be observed, associated to the appearance, or lack, of topological excitations \cite{Sanvitto2011,Amo2011}. Here we focus on
the turbulent regime where pairs of vortices nucleate at the edge of
the defect and follow the superfluid stream. In Fig. \ref{fig:panels} panel O, we show a snapshot of the polariton density when no support is present (similar setup as in Ref. \cite{Pigeon2010}). The pump with peak intensity corresponding to the black dot labeled P in Fig. \ref{fig:hyst}, is localized
in the upper part where the figure colorscale is saturated;
below it, we can distinguish the photonic defect. The other panels of
Fig. \ref{fig:panels} correspond to a sustained propagation using a
support field continuous and homogeneous in space whose intensity increases from panels A to F as noted above each panel (the labels refer to Fig. \ref{fig:hyst}). In Fig. \ref{fig:panels2} we show the phase profil of the fluid corresponding to the density profile shown in Fig. \ref{fig:panels}.
Firstly, we see, as described in Fig. \ref{fig:with_without}, that
the presence of the support allows for a superfluid propagation
over more than $100\,\unit{\mu m}$ with a constant density. This is in strong contrast with the results shown in panels O and observed in Refs. \cite{Pigeon2010,Amo2011,Nardin2011,Sanvitto2011} where the density was strongly decreasing along the propagation, modifying significantly the fluid properties. Thanks to the infinitely-extended constant support field we obtain a large and homogeneous superfluid.

\begin{figure}
\centering
\includegraphics[width=0.55\textwidth]{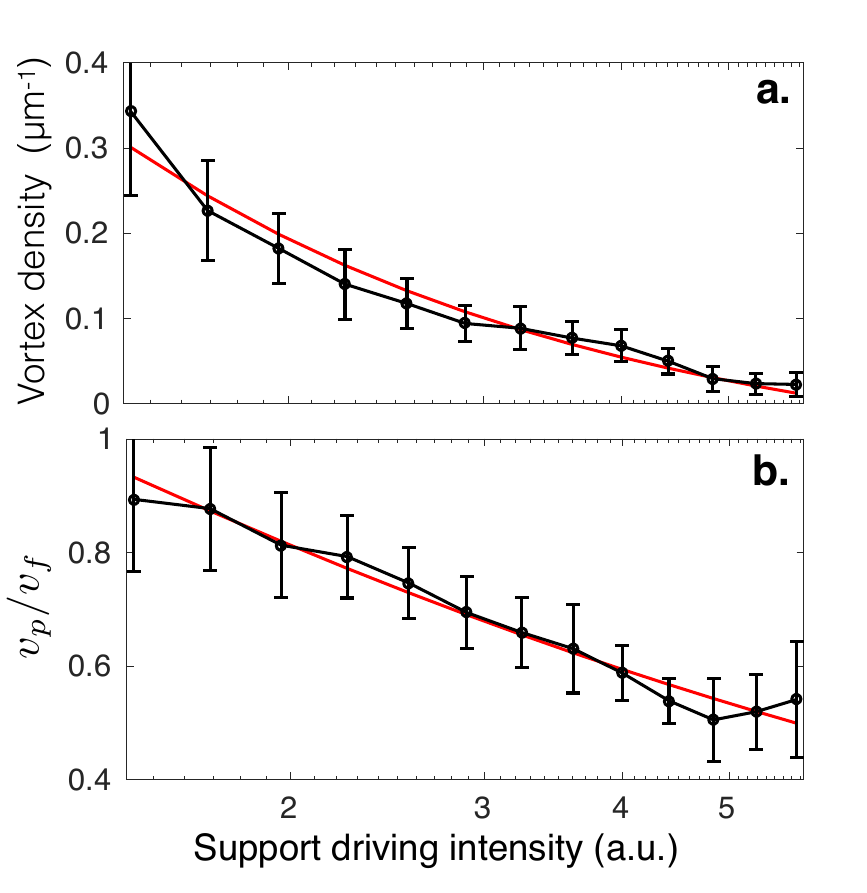}
\caption{Impact of the support field on the vortex stream. The
top panel (a.) represents the density of vortices in the stream and
the lower panel (b.) the vortex speed $v_p$ (normalized with respect
to the fluid flowing speed $v_f$). In each panel the black error bars correspond to statistical data extracted for simulations identical
to those shown in Fig. \ref{fig:panels} and in the video \cite{SupMat}.
The red curves are fits to the data with a linear dependance with the support field amplitude in (a.) and with the support field intensity in (b.). Other parameters are identical to those in Fig.
\ref{fig:with_without}. \label{fig:v_rate}}
\end{figure}

Moreover, the support provides other original effects, especially regarding its interaction with the generated vortex pairs. From panel A to F we
clearly see a vortex stream starting from the potential barrier and
following the flow, whose characteristic changes as the support
field increases. Vortices and antivortices (with opposite circulation) are
marked, respectively, with red stars and blue circles and correspond to a small density point (visible in Fig. \ref{fig:panels}) and to a phase singularity (visible in Fig. \ref{fig:panels2}). As reported in the upper panel of Fig. \ref{fig:v_rate}, the vortex density\footnote{number of vortices per micrometers along the propagation direction.} in the stream reduces as the support increases, until vortices no longer propagate. This happens even if we are still within the bistable region
(see panel E with $I_-<I_s<I_+$). The vortex density scales inversely with
the intensity of the support field. 
In Supplementary Material \cite{SupMat} 
a video presents the dynamical evolution of the polariton density over $0.5\,\unit{ns}$.
They show that for strong support, the emission of vortices still takes place, but instead of generating vortex pairs that propagate along the flow, they are forced to recombine near the potential
barrier. This effect leads to a vortex stream with constant density far away from the potential barrier not directly connected to the emission rate. The origin of this effect is not yet clearly understood, but attributed to the dual nature of the fluid, partly excitonic and partly photonic, which is, in the shadow of a purely photonic defect, is strongly modified.
Moreover, as clearly visible in video \cite{SupMat}, the vortex speed $v_p$ decreases when the support increases, as reported in the lower panel of Fig. \ref{fig:v_rate}.
Their speed is reduced by $\sim 50\%$ for increasing support by a factor of 10. For a support intensity
close to the lower limit of the bistability $I_{-}$, the vortices
flow at almost the same speed as the superfluid. Interestingly the vortex speed scales with the inverse of the
amplitude of the driving field and not with its intensity pointing out the coherent nature of this effect. 
Instead of pushing the vortices accordingly to the momentum direction imposed
by the support field, the support field slows them down; it acts as
a friction force on vortex pairs. 
Thus, by controlling the amplitude
supporting to the polariton superfluid propagation, we can control the properties of
the vortex streams. 

\begin{figure}[t!]
\centering
\includegraphics[width=0.55\textwidth]{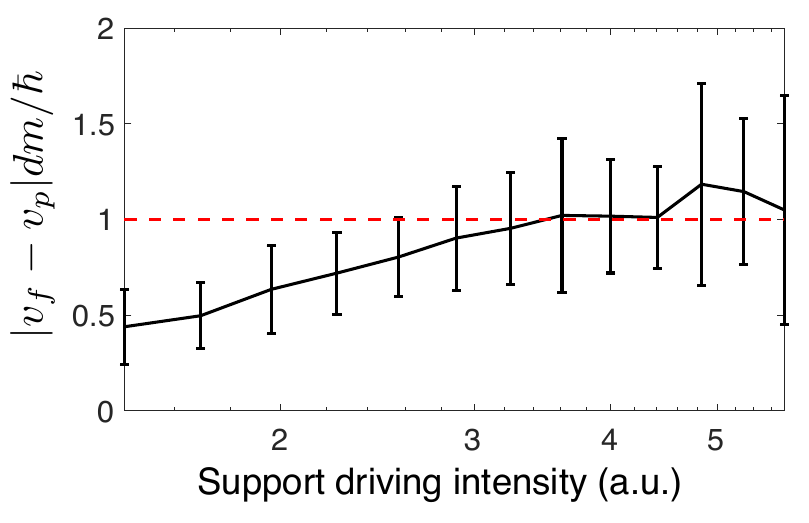}
\caption{Product of vortex speed $v_p$ with pair separation $d$ and the effective mass $m$ as a function
of the support intensity. The error bars correspond to data
extracted from the simulations and the red dashed line corresponds to expected results. \label{fig:vd}}
\end{figure}

In the present case the propagation speed of the vortex can not be understood considering single vortex interaction with the support field. The propagating entities are vortex pairs characterised by a separation distance $d$ which reduces as the support intensity increases.
The propagation of vortex pairs in irrotational liquids is a well-understood phenomenon \cite{lamb1945hydrodynamics}.
In conventional fluids, the relative speed of vortex--antivortex pairs with respect to the fluid speed evolves as 
\begin{equation}
\vert v_f-v_p \vert=\frac{\kappa}{2\pi d}
\end{equation}
with $\kappa$ the vortex circulation. In superfluids the circulation
around the vortex core is quantized as $\kappa=2\pi\hbar/m$ \cite{pitaevskii2016bose}.
Consequently the product of the pair separation $d$ by the vortex
speed $v_p$ should be equal to a constant $\hbar/m$. We report in Fig. \ref{fig:vd}
this value (red dashed line), which is independent of the support intensity,
together with the product $| v_f-v_p | d$ gathered from
the simulations (black error bars). Fig. \ref{fig:vd} clearly shows
that both results coincide for a strong support intensity. Nevertheless for a weak
support intensity (from $I_{-}$ to $I/I_{+}\approx0.15$) the deviation is
significant. The increasing uncertainty observed in Fig. \ref {fig:vd} is a direct consequence of the small number of vortices propagating for a strong support. This out-of-equilibrium superfluid exhibits interesting hydrodynamic properties that deviate significantly from standard conservative fluids. This is attributed to the driven-dissipative nature of the fluid and open the question about how far from standard hydrodynamics fluids of light stands. The resonant driving provided by the support present here is small enough to allow formation of topological excitations but still modify the properties of the fluid beyond standard hydrodynamics. This points out the need of carefully analysis and further studies in order to properly compare driven-dissipative superfluid to equilibrium one.

Notice that the fluid velocity, linked to the driving fields (support and pump) in-plane wave vectors $\mathbf{k}_p$, is the same from panel A and E of Figs.~\ref{fig:panels} and \ref{fig:panels2}, whereas the density increases by less than $10\%$. The intensity of the pump field is kept constant and corresponds to $I_r(0)/I_+>2$. For each panel the vortex propagates in a superfluid such a $v_f<c_s$. 
However wavefronts of Cherenkov type, are present near the potential barrier (Fig. \ref{fig:panels}). They
appear because the density profile close to the pump is not monotonic
(see lower panel in Fig. \ref{fig:with_without}), leading to a local supersonic regime characterized by Cherenkov wavefronts.
Upon passing the potential barrier the density stabilizes to be in a superfluid regime, and accordingly
the Cherenkov wavefront fades away \cite{Carusotto2006,Pigeon2010}.

\begin{figure}[t!]
\centering
\includegraphics[width=0.6\textwidth]{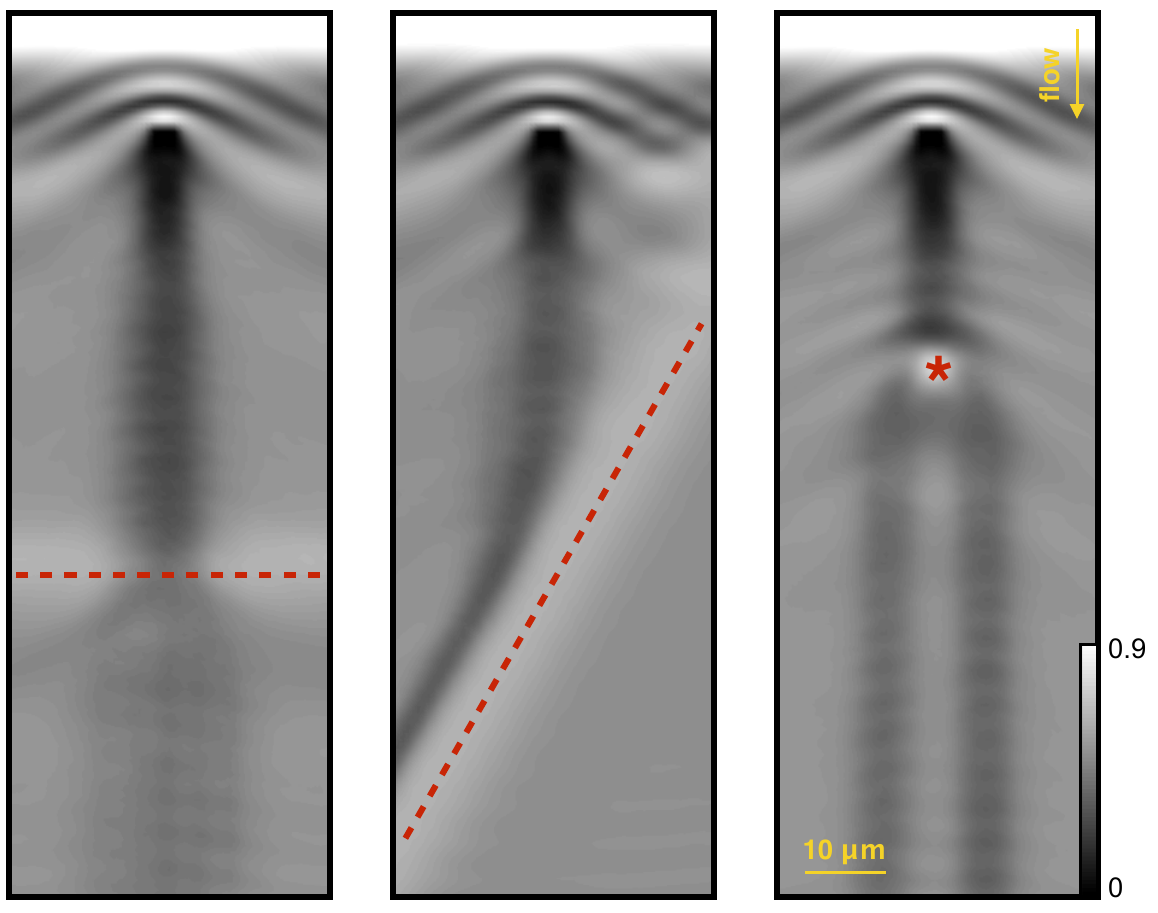}
\caption{Time integrated normalised density profile of a polariton superfluid with support
driving ($I/I_{+}=0.14$) in the presence of an extra localized driving
 highlighted by the red dashed lines and star. The
extra driving in the left and middle panels, is an asymmetric gaussian beam with a width $\sigma=2\unit{\mu m}$ along the direction normal to the dashed line and a infinite width in the normal direction. The red star in the right panel indicate the position of a symmetric gaussian beam of width $\sigma=2\unit{\mu m}$. 
The time integration is taken over $0.5\,\unit{ns}$. Parameters
correspond to Fig. \ref{fig:with_without} \label{fig:path}}
\end{figure}

When the intensity of the support is high enough, the phase of the fluid is locked to the support phase. This effect of phase locking can be used to manipulate
the propagation path of the vortex pairs. In Fig. \ref{fig:path}
we represent the polaritonic density averaged in time. Upon averaging
over time, the vortex stream appears as a low-density line, where
the local decrease of the density is directly proportional to the
vortex density.  To the pump and the support we add a third
localized continuous field (the control) indicated by the red dashed lines and
star in Fig. \ref{fig:path}. In the left panel we
add a localized control field normal to the flow, with a asymmetric gaussian spatial profile (infinite width along the dashed line and width of $2\unit{\mu m}$ in the normal direction) 
but otherwise identical properties to the two other driving fields. As a result
we obtain a broaden vortex stream. The dark region broadens and
become grayer. Instead if the localized control field is tilted with an
angle as in the middle panel of Fig. \ref{fig:path}, the stream is
directed along the line defined by the control. Finally, if the control
is focused into the stream (localised symmetrical gaussian beam) we can split it in two, as shown
in the right panel of Fig. \ref{fig:path}. 
Notice that the localised control barely changes the local density inducing no significant local energy blue shift.
The phase profil, upon averaging in time, presents very small modulations making them, in principle not mesurable. This is due for one part to the number of vortices too small to significantly impact the phase in average ; and in a second part to the stream always been formed by vortex-antivortex pairs, cancelling each other contribution to the time averaged phase profile.
With this localised field, we lock the phase of the superfluid, inhibiting locally the formation of phase modulations and therefore forcing the vortex stream to modify its trajectory.

To conclude, we numerically study in detail the interaction between
a support driving field and the propagation of a polariton superfluid. We
demonstrate how the propagation of such a superfluid can
be macroscopically extended for low power cost. Using this method we report that the release
of the phase locking effect induced by resonant support is such
that formation of topological excitations as vortices is possible.
We show that by modulating the support, the properties of
a vortex stream can be significantly manipulated. A thorough analysis of the properties of the propagating vortices
we also demonstrated that sustained out-of-equilibrium superfluid hydrodynamics
does not coincide with standard conservative hydrodynamics. 
Based on this unique feature, we illustrate how vortex path can be controlled.

This work paths the way to hydrodynamics study of polariton superfluids within an experimentally feasible setup, which was not achievable before. Moreover, it also sheds light on the phase locking mechanism characteristic of resonant driving and uses this same mechanism to achieve fine control of topological
excitations taking place within these fluids.

\ack
The authors would like to thank Mauro Paternostro, Gabriele De Chiara, and Andr\'e
Xuereb for their support as well as Elisabeth Giacobino, Iacopo Carusotto and Cristiano
Ciuti for fruitful discussions. This work was supported by the
John Templeton Foundation (grant ID 43467) and the French ANR (grant ACHN C-Flight and DS10 QFL).

\section*{References}
\bibliographystyle{apsrev4-1}

%


\end{document}